\documentclass[reqno,draft]{amsproc}
\usepackage{amsmath,amsthm,amssymb}

\newcommand{\cprime}{\/{\mathsurround=0pt$'$}}
\newcommand{\R}{\mathbb{R}}
\newcommand{\ph}{\varphi}
\renewcommand{\o}{\mathrm{o}}
\renewcommand{\d}{\partial}

\newtheorem{theorem}{Theorem}

\theoremstyle{definition}
\newtheorem*{example}{Example}

\theoremstyle{remark}
\newtheorem*{remark}{Remark}

\begin{document}
\title[Conservation laws of some evolution equations]
{Conservation laws of generalized higher Burgers \\
and linear evolution equations}
\author[Sergei Igonin]{Sergei Igonin}
\address{University of Twente, Faculty of Mathematical Sciences,
P.O. Box 217, 7500 AE Enschede, the Netherlands}
\email{igonin@mccme.ru}
%\subjclass{58F07, 58G37, 58H15, 58F37}
\begin{abstract}
By the Cole-Hopf transformation, with any linear evolution 
equation in $1+1$ dimensions 
a generalized Burgers equation is associated. We describe
local conservation laws of these equations. It turns out that any
generalized Burgers equation has only one conservation law, while
a linear evolution equation with constant coefficients
has an infinite number of $(x,t)$-independent conservation laws
iff the equation involves only odd order terms and, therefore, is
bi-Hamiltonian.
\end{abstract}
\maketitle

It is well known that the classical Burgers equation $v_t=v_{xx}+2v_xv$
is obtained from the linear heat equation $u_t=u_{xx}$ by the
Cole-Hopf transformation $v=u_x/u$, but it is less known
that this construction can be applied to an arbitrary linear evolution equation
\begin{equation}
  \label{lin}
 u_t=\sum_{i=0}^m a^i(x,t)u_i,
\end{equation}
where, as usual, $u_i={\d^i u}/{\d x^i}$ and
\begin{equation}
  \label{m}
  a^m(x,t)\neq 0.
\end{equation}

Indeed, denote $v=v_0=u_1/u$ and $v_k=D_x^k(v)$, then
$$
v_t=D_tD_x(\ln|u|)=
D_x(\frac{u_t}u)=D_x\Bigl(\sum_{i=0}^m a^i(x,t)(D_x+v)^i(1)\Bigl),
$$
where $D_x$ and $D_t$ are the total derivatives,
because it is easily seen that 
$$
\frac{u_i}{u}=(D_x+v)^i(1).
$$ 
The equation
\begin{equation}
  \label{burg}
  v_t=D_x\Bigl(\sum_{i=0}^m a^i(x,t)(D_x+v)^i(1)\Bigl)
\end{equation}
is called the {\it generalized Burgers equation} associated with
linear equation \eqref{lin}.
This is an example of nonlinear $\mathrm{C}$-integrable equations in
the sense of Calogero \cite{C}.
\begin{example}
Consider linear equations with constant coefficients
\begin{equation}
  \label{lin1}
 u_t=\sum_{i=0}^m a^i u_i,\ a_m\neq 0.
\end{equation}
Clearly, flows \eqref{lin1} mutually
commute for arbitrary $a_i\in\R$, hence flows \eqref{burg} with
constant coefficients $a^i(x,t)=a^i$ also commute and form the
integrable {\it Burgers hierarchy} (see, for example, \cite{burg}).
It passes the Painlev\'e test \cite{P}, 
but, as it follows from Theorem \ref{clB} below, any generalized Burgers
equation has only one local conservation law. Note that this seems to
be the only nonlinear integrable hierarchy that allows complete
description of conservation laws for all its members.
\end{example}

In the present article we study conservation laws of equations
\eqref{lin}, \eqref{burg}. This question is trivial in the case of
evolution equations of order $1$, and from now on we assume $m\ge 2$.

Recall \cite{rb,olver} that a smooth function
$\ph(x,t,u,u_1,\dots,u_k)$ is called a {\it conserved density} of an
equation
\begin{equation}
  \label{evol}
 u_t=F(x,t,u,u_1,\dots,u_m)
\end{equation}
if $D_t(\ph)=D_x(\psi(x,t,u,u_1,\dots,u_l))$ for some function $\psi$.
Two conserved densities $\ph_1,\,\ph_2$ are said to be {\it equaivalent}
if $\ph_1-\ph_2=D_x(\psi)$, and conserved densities of the form
$D_x(\psi)$ are called {\it trivial}.
Equivalence classes of conserved densities are called
{\it conservation laws}.
\begin{example}
The function $v$ is a nontrivial conserved density for \eqref{burg}.
\end{example}
Let $\ph$ be a conserved density for \eqref{evol}, then its
{\it characteristic} \cite{olver} (also called the
{\it generating  function} \cite{rb,vin}) is the variational
derivative of $\ph$
\begin{equation}
  \label{ch}
  \xi=\frac{\delta\ph}{\delta u}=\sum_{i\ge 0}(-1)^iD_x^i(\frac{\d\ph}{\d u_i}).
\end{equation}
The characteristic satisfies the equation
\begin{equation}
  \label{che_g}
  -D_t(\xi)=\sum_{i=0}^m (-1)^iD_x^i(\frac{\d F}{\d u_i} \xi)
\end{equation}
and is nonzero if and only if the density $\ph$ is nontrivial, i.e.,
equivalent conserved densities have the same characteristic.
The homological interpretation of these concepts can be found in
\cite{rb,vin}.

For a function $h(x,t,u,u_1,\dots)$ the maximal integer $k$ such that
${\d h}/{\d u_k}\neq 0$ will be called the {\it order} of $h$ and
denoted by $\o(h)$. If ${\d h}/{\d u_k}=0$ for all $k>0$,
we set $\o(h)=0$.
\begin{theorem}
\label{xi}
The characteristic $\xi$ of any conserved density for \eqref{lin} has the
form
\begin{equation}
  \label{ch_lin}
  \xi=a(x,t)+\sum_{i}b^i(x,t)u_i
\end{equation}
such that the operator $\sum_i b^i(x,t)D_x^i$ is self-adjoint,
i.e.,
\begin{equation}
  \label{sa}
  \sum_i b^i(x,t)D_x^i(\psi)=\sum_i (-1)^iD_x^i(b^i(x,t)\psi).
\end{equation}
for any smooth function $\psi=\psi(x,t,u,u_1,\dots)$.
\end{theorem}
\begin{proof}
We must prove
\begin{equation}
  \label{sol}
  \frac{\d^2\xi}{\d u_i\d u_j}=0,\ \ \forall\,i,\,j\ge 0,
\end{equation}
then \eqref{sa} follows from the properties of variational derivatives
\cite{rb,olver}.
If \eqref{sol} is not true then there exist $i_0\ge j_0\ge 0$ such
that
\begin{equation}
  \label{as}
  \frac{\d^2\xi}{\d u_{i_0}\d u_{j_0}}\neq 0
\end{equation}
and
\begin{equation}
  \label{sol1}
  \frac{\d^2\xi}{\d u_i\d u_j}=0,\ \ \forall\,i,j: i>i_0,\,i+j\ge i_0+j_0.
\end{equation}
For \eqref{lin} equation \eqref{che_g} reads
\begin{equation}
  \label{che}
  -D_t(\xi)=\sum_{i=0}^m (-1)^iD_x^i(a^i(x,t)\xi).
\end{equation}
Differentiating \eqref{che} with respect to $u_{\o(\xi)+m}$,
we get
$$
-a^m(x,t)\frac{\d\xi}{\d u_{\o(\xi)}}=
(-1)^m \cdot a^m(x,t)\frac{\d\xi}{\d u_{\o(\xi)}}.
$$
Combining this identity with \eqref{m}, we see that $m$ is odd.
Now differentiating \eqref{che} with respect to $u_{i_0+m-1},\,u_{j_0+1}$ and
taking into account \eqref{sol1}, one obtains
$$
-a^m(x,t)\frac{\d^2\xi}{\d u_{i_0-1}\d u_{j_0+1}}
=-a^m(x,t)\frac{\d^2\xi}{\d u_{i_0-1}\d u_{j_0+1}}
-m\cdot a^m(x,t)\frac{\d^2\xi}{\d u_{i_0}\d u_{j_0}},
$$
which contradicts to our assumption \eqref{as}; hence \eqref{sol} is true.
\end{proof}
\begin{theorem}
\label{clB}
Any conserved density of a generalized Burgers equation \eqref{burg}
is equivalent to $Cv$ for some $C\in\R$, i.e., the space of
conservation laws is one-dimensional.
\end{theorem}
\begin{remark}
For the classical Burgers equation this fact is well known and follows
from the general observation \cite{rb,olver} that for any even order evolution
equation there is an upper bound on the order of the characteristics
of conservation laws. However, these arguments do not work for
generalized Burgers equations of arbitrary order considered in this theorem.
\end{remark}
\begin{proof}
For a conservation law $\Omega$ of \eqref{burg} consider a
conserved density $\ph\in\Omega$ of minimal order $k$.
Replacing $v_k$ by $D_x^k(u_1/u)$ in $\ph$, we obtain
a conserved density $\tilde \ph$ for \eqref{lin}.
By the construction, $\tilde \ph$ is invariant under the symmetry
$u_k\mapsto\lambda u_k$ of \eqref{lin}, $k\ge 0,\,\lambda\in\R$. Then
from \eqref{ch} for the characteristic $\xi$ of $\tilde \ph$ one has
\begin{equation}
  \label{sym}
  \xi(x,t,\lambda u,\dots,\lambda u_{2k+2})=
  \lambda^{-1}\xi(x,t,u,\dots,u_{2k+2}),\ \ \forall\,\lambda\neq 0.
\end{equation}
Combining \eqref{sym} with \eqref{ch_lin}, we obtain $\xi=0$, that is,
\begin{equation}
  \label{triv}
  \tilde\ph=D_x\psi(x,t,u,\dots,u_{k})
\end{equation}
for some function $\psi$.

From \eqref{triv} we have ${\d^2\ph}/{\d u_{k+1}^2}=0$, therefore,
${\d^2\ph}/{\d v_k^2}=0$.
This implies $k=0$, because if $k>0$, one easily constructs
an equivalent conserved density $\ph'$ with $\o(\ph')<k=\o(\ph)$,
which contradicts to our assumption that $\ph\in\Omega$ is of minimal order.
Thus $\ph=l(x,t)v+n(x,t)$, $\tilde\ph=l(x,t)u_1/u+n(x,t)$, and from
\eqref{triv} we obtain that $l(x,t)=l(t)$ does not depend on $x$.
Finally, from the condition that $\ph$ is a conserved density it follows
that $l(t)$ is actually a constant.
\end{proof}

In contrast to generalized Burgers equations, for a linear equation
\eqref{lin} the space of conservation laws may be
infinite-dimensional.
\begin{theorem}
For a linear evolution equation with constant coefficients \eqref{lin1}
the space $\mathcal{C}\!\mathcal{L}$ of $(x,t)$-independent
conservation laws is described as follows.
\begin{enumerate}
\item If $a^0\neq 0$ then $\mathcal{C}\!\mathcal{L}=0$.
\item If $a^0=0$, but $a^{2j}\neq 0$ for some $j\in\mathbb{N}$ then
$\mathcal{C}\!\mathcal{L}$ is one-dimensional and generated by the
conserved density $u$.
\item If all the even coefficients $a^{2j}$ vanish,
$\mathcal{C}\!\mathcal{L}$ is infinite-dimensional and generated by
the conserved densities
\begin{equation}
  \label{cl}
  u,\ u_k^2,\ \ k\ge 0.
\end{equation}
\end{enumerate}
\end{theorem}
\begin{proof}
Let $\ph=\ph(u,\dots,u_k)$ be an $(x,t)$-independent conserved density.
From \eqref{ch_lin} and \eqref{sa} it follows that the characteristic
$\xi$ of any $(x,t)$-independent conserved density has the form
\begin{equation}
  \label{ch_const}
  \xi=c+\sum_{j} c^j u_{2j},\ \  c,\,c^j\in\R.
\end{equation}
Equation \eqref{che} reads
\begin{equation}
  \label{che1}
  -\sum_{j}\sum_{i=0}^m a^ic^ju_{i+2j}=
  \sum_{j}\sum_{i=0}^m (-1)^{i}a^ic^ju_{i+2j}+a^0c.
\end{equation}
Clearly, if there is a nonzero even coefficient $a^{2j}\neq
0,\,j\in\mathbb{N},$ then \eqref{che1} implies $\xi=c\in\R$,
while if $a^0\neq 0$ then $\xi=0$. On the other hand, if $a^{2j}=0$
for all $j$ then \eqref{che1} is valid for any constants $c,\,c^j$.
It is easily seen that if $a^0=0$ then $cu$ is a conserved density
with the characteristic $\xi=c$, while if $a^{2j}=0$ for all $j$ then
the function $\ph=u_j^2$
is a conserved density with characteristic $\xi=(-1)^j 2u_{2j}$.
Hence every function \eqref{ch_const} satisfying \eqref{che1} is indeed
a characteristic of a linear combination of the above described
conserved densities, which thus span the whole space of conservation laws.
\end{proof}
%\begin{remark}
%If $a^{2j}=0$ for all $j$, each solution of equation \eqref{che_g} with
%$F=\sum_i a^i u_i$ is a (generalized) symmetry of \eqref{lin1}. In
%particular, characteristics \eqref{ch_const} of conservation laws are symmetries.
%\end{remark}
%\begin{remark}
Conservation laws \eqref{cl} of a linear equation   
\begin{equation}
  \label{lin3}
  u_t=\sum_{j\ge 0}a^{2j+1}u_{2j+1},\ a^{2j+1}\in\mathbb{R},
\end{equation}
can be derived as follows. Performing the Galilean transformation 
$$
x\mapsto x+a^1t,\ t\mapsto t,\ u\mapsto u,
$$ 
we obtain an equation of the form \eqref{lin3} with $a^1=0$. Such an
equation possesses a bi-Hamiltonian structure \cite{H}
\begin{equation}
  \label{bH}
  \begin{aligned}
  u_t & = D_x\Bigl(\frac{\delta}{\delta u}\sum_j(-1)^j\frac12 a^{2j+1}u_{j}^2\Bigl),\\   
  u_t & =
   D_x^3\Bigl(\frac{\delta}{\delta u}\sum_{j>0}(-1)^{j-1}\frac12 a^{2j+1}u_{j-1}^2\Bigl).    
  \end{aligned}
\end{equation}
It is easily seen that conservation laws \eqref{cl} are obtained
from this by the Lenard scheme \cite{L,olver} and, therefore,
are in involution with respect to the both corresponding Poisson brackets.
According to \cite{H}, using \eqref{bH} and the Cole-Hopf
transfomation, one can also construct a nonlocal bi-Hamiltonian structure
for the generalized Burgers equation associated with \eqref{lin3}.  
\section*{Acknowledgments}
The author would like to thank Prof.~I.~S.~Krasil{\cprime}shchik and
Prof.~R.~Martini for useful remarks.

%%%%%%%%%%%%%%%%%%%%%%%%%%%%%%%%%%%%%%%%%%%%%%%%%%

\end{document}